\documentclass[sigconf,nonacm]{acmart}

\usepackage{subfigure}
\usepackage[shortlabels]{enumitem}

\AtBeginDocument{%
  \providecommand\BibTeX{{%
    \normalfont B\kern-0.5em{\scshape i\kern-0.25em b}\kern-0.8em\TeX}}}

\copyrightyear{2021} 
\acmYear{2021} 
\setcopyright{acmlicensed}
\acmConference[CHIIR '21]{Proceedings of the 2021 ACM SIGIR Conference on Human Information Interaction and Retrieval}{March 14--19, 2021}{Canberra, ACT, Australia}
\acmBooktitle{Proceedings of the 2021 ACM SIGIR Conference on Human Information Interaction and Retrieval (CHIIR '21), March 14--19, 2021, Canberra, ACT, Australia}
\acmPrice{15.00}
\acmDOI{10.1145/3406522.3446035}
\acmISBN{978-1-4503-8055-3/21/03}

\begin{document}
\fancyhead{} 

\title{Starting Conversations with Search Engines - Interfaces that Elicit Natural Language Queries}

\author{\mbox{Andrea Papenmeier, Dagmar Kern, Daniel Hienert}}
\email{firstname.lastname@gesis.org}
\affiliation{%
  \institution{GESIS – Leibniz Institute for the Social Sciences}
  \city{Cologne}
  \country{Germany}
}

\author{Alfred Sliwa, Ahmet Aker, Norbert Fuhr}
\email{firstname.lastname@uni-due.de}
\affiliation{%
  \institution{University of Duisburg-Essen}
  \city{Duisburg}
  \country{Germany}
}

\renewcommand{\shortauthors}{Papenmeier et al.}

\begin{abstract}
   Search systems on the Web rely on user input to generate relevant results. Since early information retrieval systems, users are trained to issue keyword searches and adapt to the language of the system. Recent research has shown that users often withhold detailed information about their initial information need, although they are able to express it in natural language. We therefore conduct a user study (N = 139) to investigate how four different design variants of search interfaces can encourage the user to reveal more information.
   Our results show that a chatbot-inspired search interface can increase the number of mentioned product attributes by 84\% and promote natural language formulations by 139\% in comparison to a standard search bar interface.
\end{abstract}



\maketitle

\section{Introduction}

When searching the Web with search engines, users have to communicate what they are looking for in a machine-interpretable manner. Search engines focusing on products often offer additional possibilities to navigate, such as filters or facets. However, research on product search engines shows that only a fraction of the actual information need is present in the initial search query \cite{papenmeier2020modern}. A reason for this  could be the vocabulary problem \cite{furnas1987vocabulary}: The structured data in the databases use a different vocabulary than the natural language of users. Hence, the user has to adapt to the system's language to maximise search success.

With voice assistants and voice search being on the rise, search engines need to process longer and more complex inputs. When equipped with appropriate capabilities to process such input, search engine performance can even be improved \cite{bendersky2008discovering}. This opens up new opportunities: If users express their information need through natural language and disclose more information, search performance can be improved. However, keyword search is deeply ingrained in the users' minds \cite{kammerer2012children}.
\begin{figure}[tb]
    \includegraphics[width=0.85\linewidth]{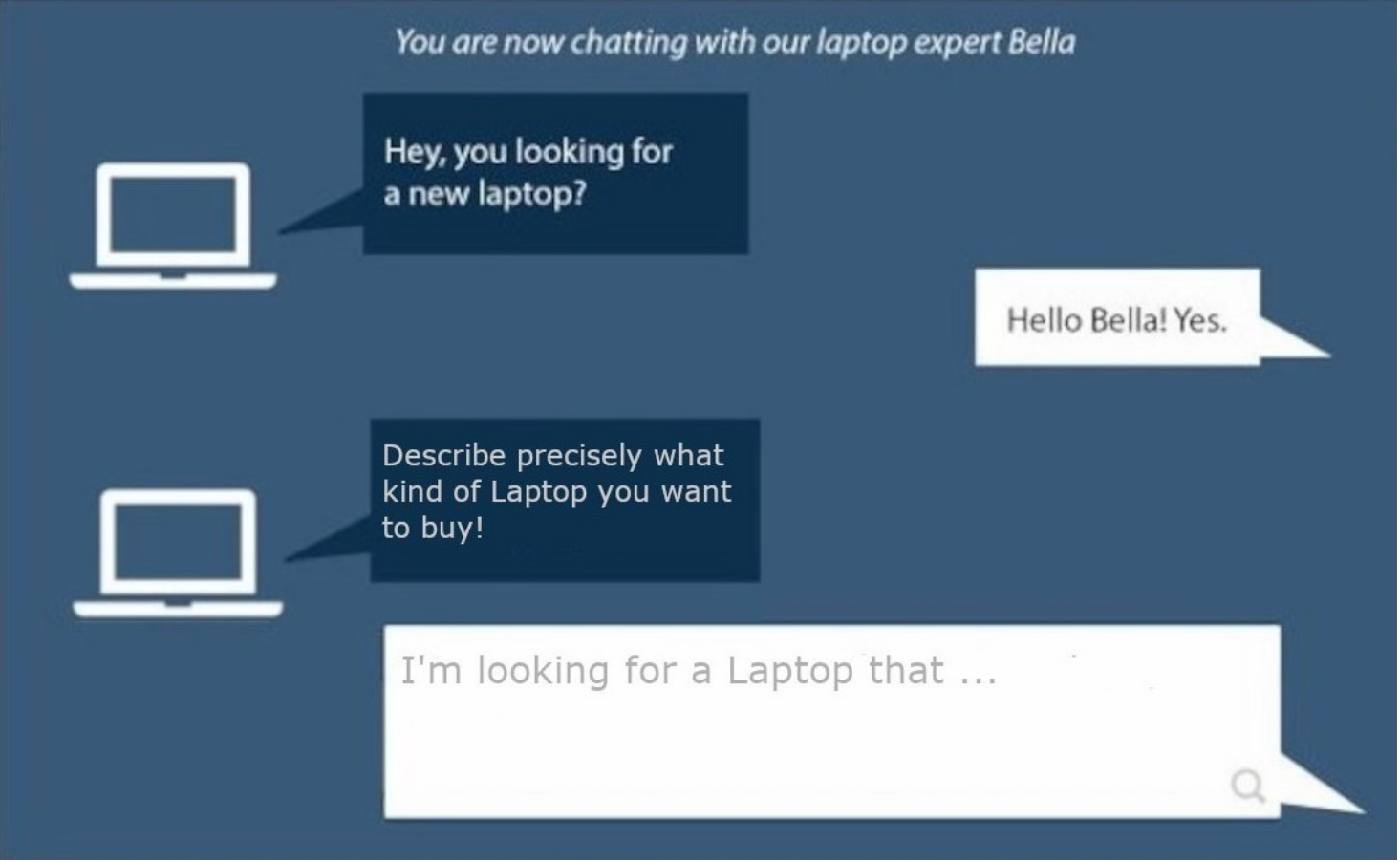}
    \caption{Final design of the chatbot interface.}
    \label{fig:interface_chatbot}
\end{figure}

In our research, we investigate how an interface can trigger users to give more information on their needs directly at the first interaction in a product search scenario. We design a user study to evaluate and compare four interfaces concerning the differences in query length, information content and formulation characteristics. Our key findings show that a chatbot-inspired interface (see Figure \ref{fig:interface_chatbot}) succeeds in eliciting more information about the user's information need than traditional search bar interfaces, while also triggering a more natural query language.

\section{Related Work}
Although research showed that longer queries could produce better results for information seeking tasks, e.g. \cite{belkin2002Rutgers,Belkin2003}, people usually tend to use short search queries \cite{Bailey2010}. There are many approaches to support users in finding relevant information, e.g. through facets, recommendations, implicit and explicit user feedback. However, only few works have tried to motivate users to type in more query terms and thus provide more detailed information about their initial information need. Belkin at al., for example, showed that a query-entry box with several lines led to longer queries than a line mode search bar ~\cite{belkin2002Rutgers} and that query lengths were significantly longer when the query box was labelled with ``Information problem description (the more you say, the better the results are likely to be)'' than when it was labelled with ``Query terms'' \cite{Belkin2003}. Furthermore, they found that longer queries significantly increase searchers' satisfaction \cite{Belkin2003}. In contrast, Agapie et al.~\cite{Agapie2013} found that telling users that longer queries deliver better search results does not influence query length. However, they showed that using a coloured halo around the search bar motivates searchers to provide significantly more query terms in a complex Web search scenario. Hiemstra et al.~\cite{Hiemstra2017} evaluated the proposed halo effect in a website search system in a 50-day A/B test (N = 3506) but could not confirm the positive impact on query length. They conclude that this approach might be sensitive to the search task and search context. Kelly and Fu~\cite{KELLY2007} show that additional information (domain knowledge, the information need, and search motivation) help to increase the retrieval performance. Likewise, Bendersky, Croft and Bruce \cite{bendersky2008discovering} propose a machine learning method to extract the key facts from long queries. Their system performs better on longer natural language queries as compared to shorter, keyword-like queries. 

A reliable information need elicitation is getting more critical with the increasing use of voice assistant systems. Without a graphical user interface, refining the search via facets and exploring the results and recommendation lists becomes cumbersome. Research in the context of conversational search has explored asking clarifying questions \cite{aliannejadi2019asking} or coached conversational preference elicitation \cite{radlinski-etal-2019-coached}. With one good question, Aliannejadi et al.~\cite{aliannejadi2019asking} improved the retrieval performance by over 150\%. Focusing on conversations, however, requires processing natural language (with challenges such as vague language and ambiguity \cite{balfe2004improving}), which is, so far, not supported by common product search engines.

\section{User Study}
Previous research has explored changes on the graphical user interface and in the interaction design to elicit longer search queries. User studies, so far, have not focused on the generation of natural language queries to support elicitation of the natural and complete information need, especially not in the context of product search. We close this gap by investigating the following research questions:
\begin{enumerate}[-,topsep=5pt]
    \item [\textbf{RQ 1:}] Do users reveal more about their information need when interacting with more conversation-like interfaces?
    \item [\textbf{RQ 2:}] Are users more inclined to use natural language when interacting with more conversation-like interfaces?
\end{enumerate}
To answer the research questions, we design interfaces based on cues from prior literature and evaluate them in an online study with a between-subjects design. The interface designs, questionnaire, and annotation guidelines are available online\footnote{\url{https://git.gesis.org/papenmaa/chiir21_naturallanguageinterfaces}}.

\subsection{Iterative Interface Design}
Initially, seven interfaces were designed that implemented either different search bar sizes (inspired by \cite{belkin2002Rutgers}), a dialogue-inspired speech bubble and chatbot design inspired by conversational search, or different avatars (inspired by agents in e-commerce \cite{Palopoli2006Agents}). All seven interfaces were tested in a between-subject online pilot study with 60 participants. Besides issuing a query, the participants gave qualitative feedback about their experiences. Finally, the set of interfaces was reduced to four interfaces that showed the most diversity in query formulations. Interfaces using avatars were excluded to focus on search bars and dialogue-inspired designs: an interface with a small search bar (I1), one with a big search bar (I2), one with a direct question in a speech bubble (I3), and a chatbot (I4, see Figure \ref{fig:interface_chatbot}). As placeholders (or the absence of placeholders) impact the query formulation, we use the exact same placeholder text for all interfaces (``I'm looking for a laptop that...''). Hence, effects of the placeholder impact all conditions equally.

\subsection{Scenario and Procedure}
To evaluate the interfaces, we set up an online study on SoSci Survey\footnote{https://www.soscisurvey.de/en/index}. After giving consent, participants receive the scenario and task description (adapted from \cite{barbu2019influence,papenmeier2020modern}):
\begin{quote}
\centering
    \textit{Your laptop broke down yesterday. Today, you are searching for a new device. You decide to search online. You find the following website. Please use the search bar in the screenshot to search for your desired laptop.}
\end{quote}
For I1, I2, and I3, participants submit their query and are redirected to a dummy result page to complete the search experience. For I4, participants submit the initial query, but then receive a generic follow-up question from the chatbot, which asks them to give more details. This design is chosen to simulate a chatot-like interaction. After the second prompt, participants of I4 also continue on the dummy result page. Subsequently, three open questions about their experience are asked, as well as closed questions about the individual domain knowledge and demographic background.

\subsection{Measures and Analysis}
To investigate whether the queries differ in the informational content (\textbf{RQ 1}), we analyse the submitted search query with the following measures: (a) count of words, (b) count of key facts -- key facts being descriptions of product attributes, e.g. ``i5 processor'', ``large screen'' --, and (c) the attribute group per key fact (e.g. ``processor'', ``screen''). For analysing differences in formulations concerning natural language (\textbf{RQ 2}), we determine per query: (d) count of vague words -- ambiguous words that cannot clearly be mapped onto product attributes or values e.g. ``good'', ``decent'' --, (e) occurrences of grammatical word types (part of speech), and (f) sentence completeness (scale of 0-2, with 0 = keywords or bullet points, 1 = partial sentences, and 2 = full sentences). We use the NLTK\footnote{https://www.nltk.org} Python package for automatically retrieving (a) and (e). For extracting (b), (c), (d) and (f), we manually annotate the queries with two annotators who discuss discrepancies until a final annotation is found.

Finally, to test for significant differences between a group of independent samples (e.g. comparing conditions), we use the Kruskal-Wallis test with a pairwise Mann-Whitney U-test as post-hoc analysis, applying the Bonferroni correction to account for the multiple testing bias. For analysing correlations, we use the measure of Spearman's rho. 

\subsection{Participants}
Overall, we recruited 139 participants (57 male, 80 female, 2 diverse) on the online crowdsourcing platform Prolific\footnote{www.prolific.co}.
Participants were evenly spread over the four interfaces (N = {36, 34, 34, 35}). As the sample group per condition is rather small, we aimed for a homogeneous sample: Users had to be residents of the US, the UK, or Ireland, native English speakers, and had to be ``digital native'' adults (ages 18-40, M = 28.3, STD = 5.9).

\begin{figure*}[t]
  \centering
  \subfigure[query length (words)]{\includegraphics[width=0.3\textwidth]{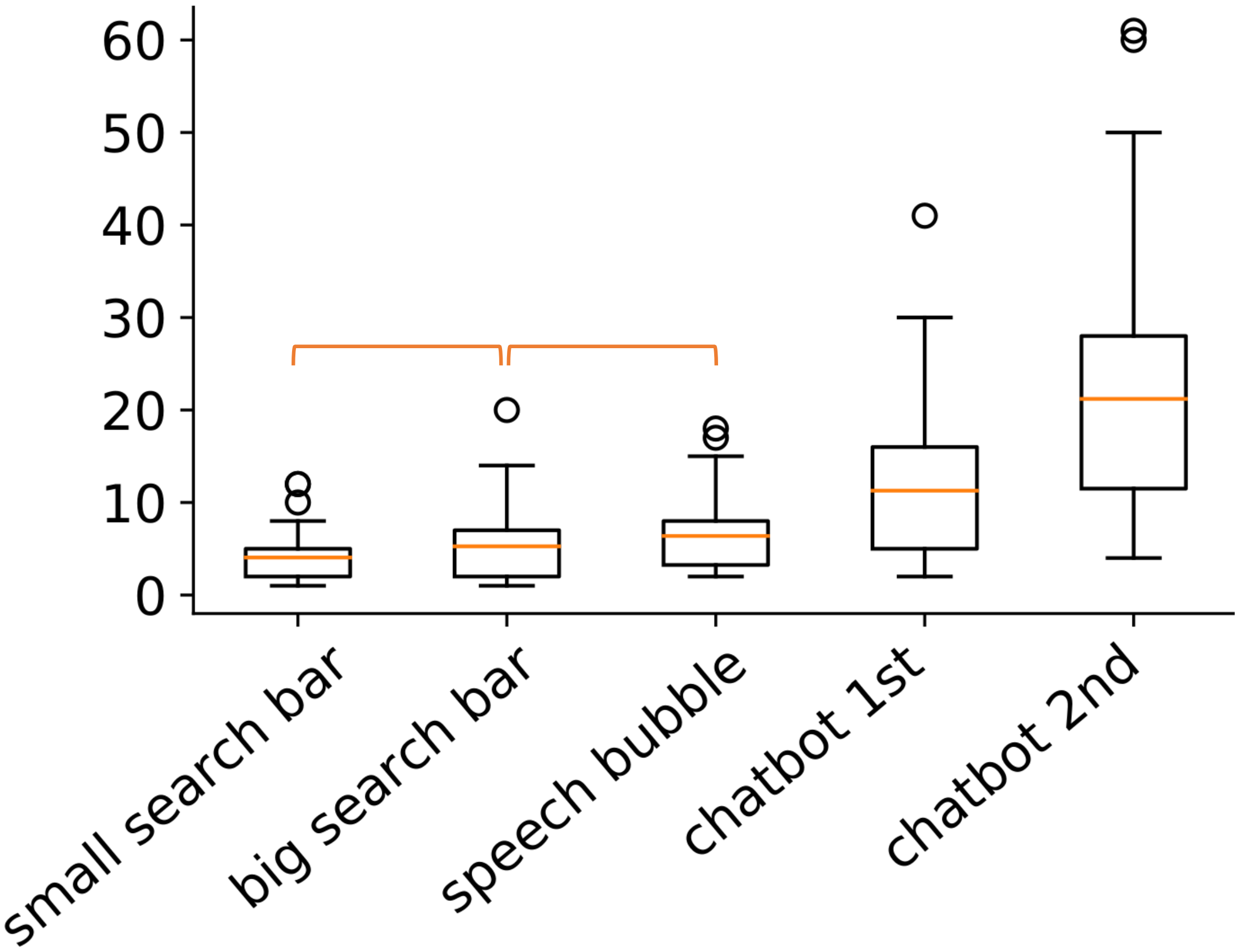}}\quad
  \subfigure[keyfact count]{\includegraphics[width=0.3\textwidth]{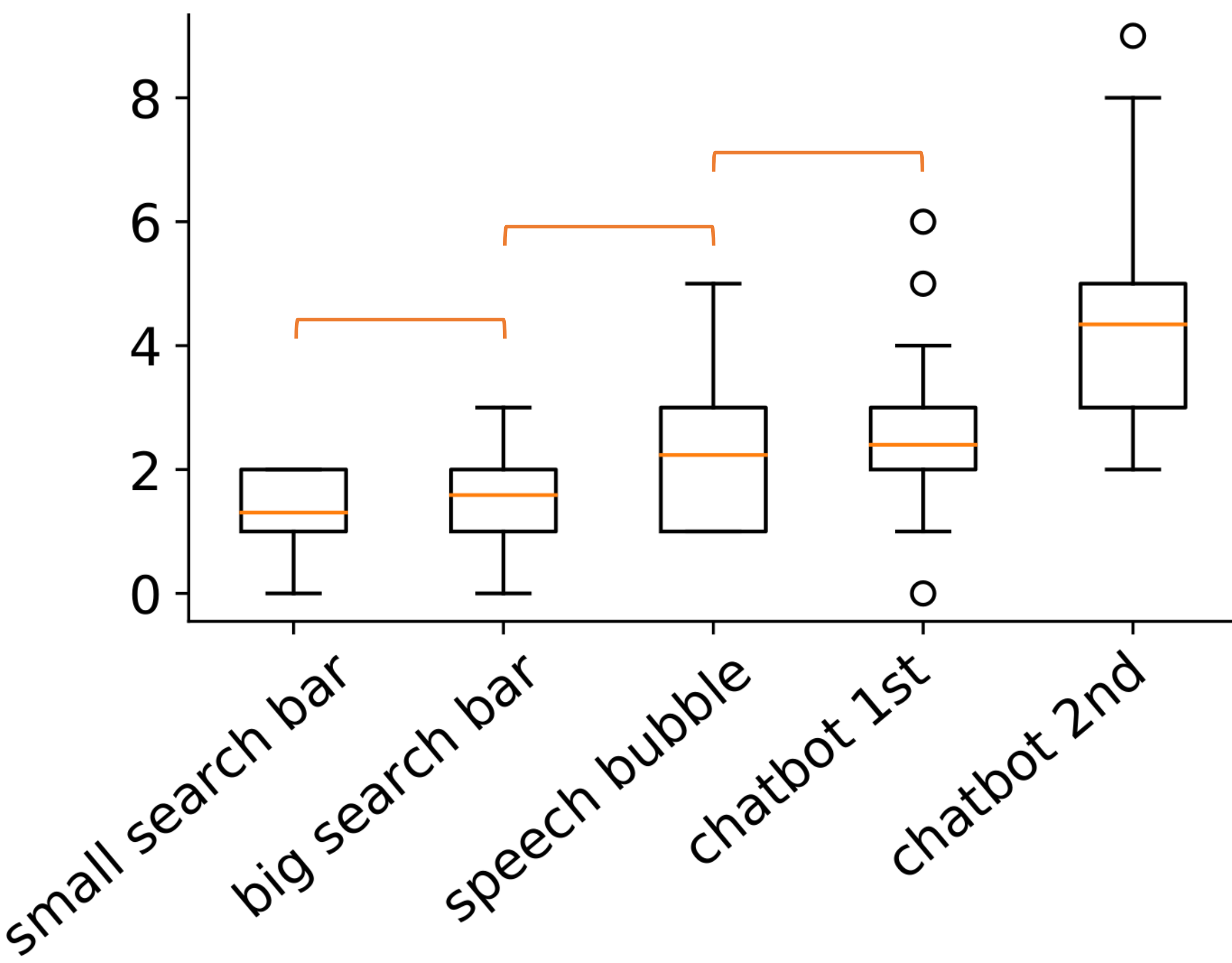}}\quad
  \subfigure[vague word count]{\includegraphics[width=0.3\textwidth]{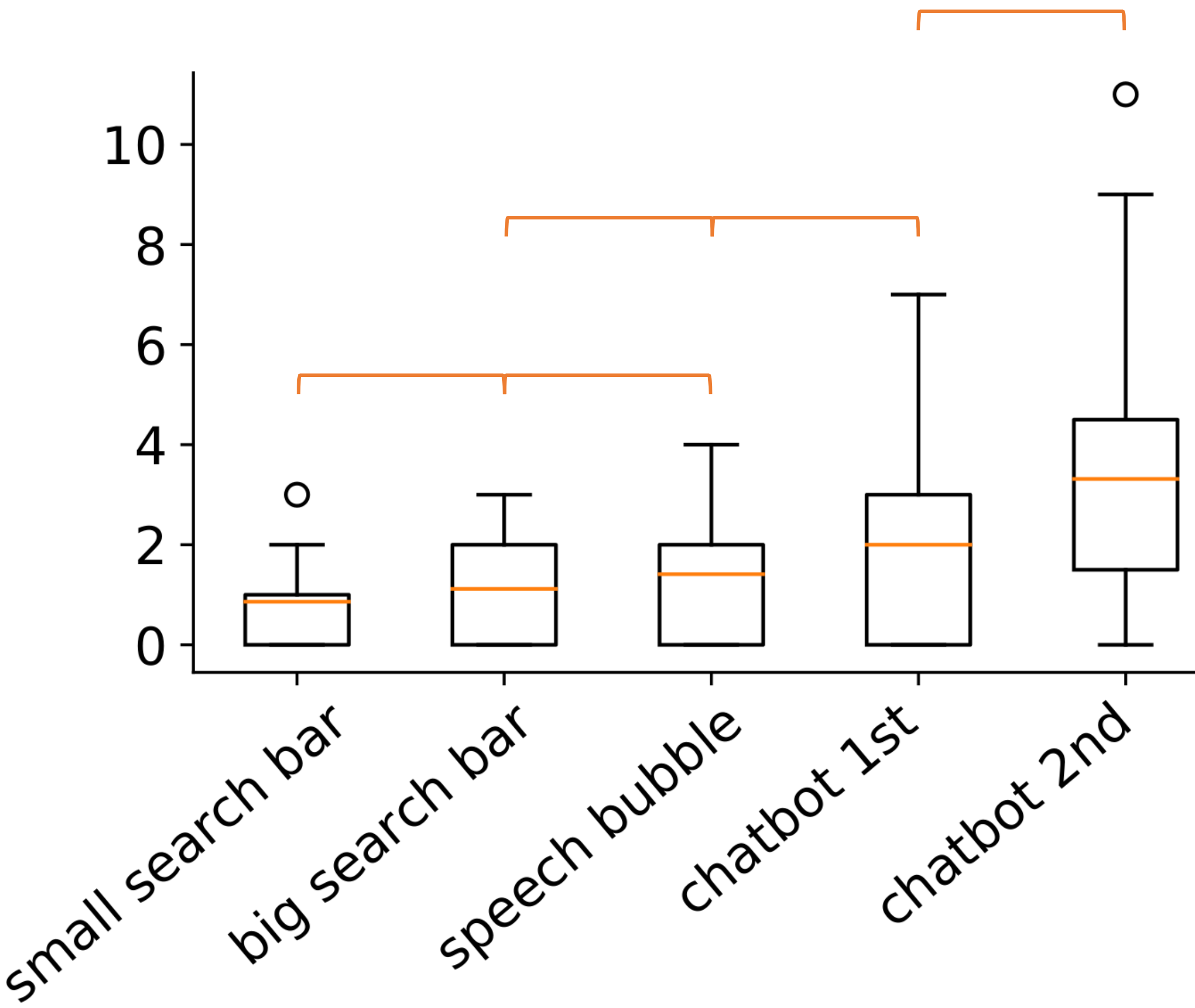}}
  \caption{Boxplots of query length (a), key fact count (b), and vague words (c). Orange brackets signify an absence of significant difference between conditions, no bracket means significant difference between conditions.}
  \label{fig:boxplots}
\end{figure*}

\section{Results}

\subsection{Information Content}
To investigate whether users reveal a different amount of information about their preferred product across conditions, we analyse the queries concerning their length, number of individual key facts, and the attribute groups. An exemplary query issued in I3 is given in the following, with key facts marked in bold:
\begin{quote}
\centering
    \textit{``I'd like something \textbf{lightweight}, \textbf{easy to use} with a \textbf{long battery life}} 
\end{quote}

The initial queries of I4 are significantly longer (M = 11, STD = 9) and contain significantly more key facts (M = 2.4, STD = 1.2) than both search bar interfaces I1 and I2 (see Figures \ref{fig:boxplots}a and \ref{fig:boxplots}b). On average, I4 leads to 84\% more key facts than I1. A generic follow-up question can further achieve a significant raise in length (M = 21, STD = 15) and key facts (M = 4.3, STD = 1.7). We found a moderately positive correlation between domain knowledge and query length (r = .41). A difference in domain knowledge could bias the results if the average domain knowledge differed across conditions. A Kruskal-Wallis test, however, did not show any difference in domain knowledge (p = .564). Although participants with more domain knowledge tend to write longer queries, there is no correlation between domain knowledge and the number of key facts in a query (r = .01). However, comparing the length and the number of key facts, we find a strong positive correlation (r = .67). 

We clustered all key facts into groups according to the product attribute they were addressing. Overall, we identified 26 unique product attributes, of which 13 appeared in queries of all conditions: brand, model, memory, graphics card, RAM, screen, battery, size, software, price, performance, quality, and purpose. In queries of I1, we found 15 unique attributes, 17 in I2, 22 in I3, 20 in I4, and 23 in I4 after the follow-up question. Attributes that were only mentioned in a dialogue-inspired condition (but not in a search bar condition) were, for example, the design of a laptop, the keyboard or the usability. This shows that the dialogue-inspired designs (I3, I4) elicited not only more key facts but also a greater variety of key facts.

\subsection{Natural Language}
Besides the information content, we aim to investigate the usage of natural language -- both concerning the interface design, as well as the information value level.

First, we investigate the sentence completeness by analysing whether participants formulated their query as a keyword search, using partial sentences, or using full sentences. In I4, 43\% of queries are grammatically complete sentences, e.g.:
\begin{quote}
\centering
    \textit{``I want a laptop that is designed by apple for business purposes. Between \pounds700-\pounds1000 and rose gold''} 
\end{quote}
In I1, I2, and I3, only 18\%, 12\%, and 11\% of the participants used complete sentences. Compared to I1, the usage of complete sentences increased significantly by 139\% in I4. But also compared to I2 and I3, I4 brings about significantly more complete sentences. Keyword search shows the inverse trend, e.g.:
\begin{quote}
\centering
    \textit{``windows laptop 8gb RAM''} 
\end{quote}
Only 3\% of the participants in I4 formulated their query in such a bullet point form, compared to 61\% in I1 and 53\% in I2. The results show a weakly positive correlation between sentence completeness and the number of key facts mentioned in a query (r = .20): The more complete the sentences were formulated, the more key facts were mentioned in the query.

\begin{figure}[t]
    \includegraphics[width=\linewidth]{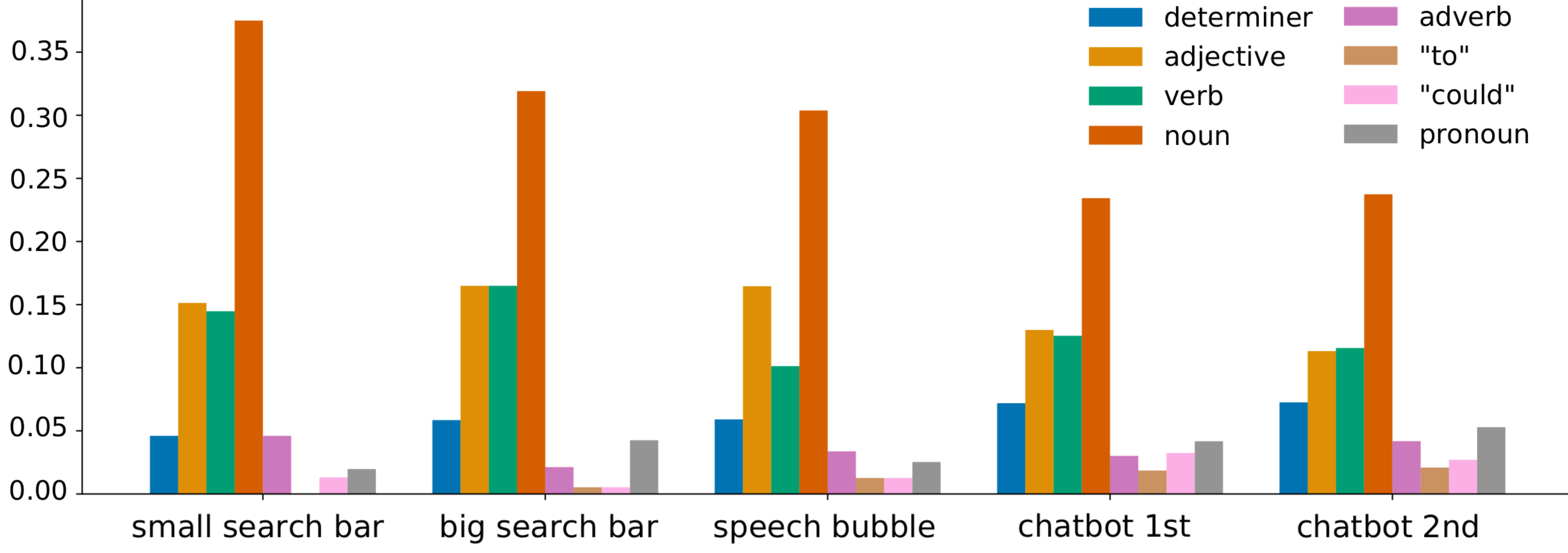}
    \caption{Distribution of most frequent parts of speech within each condition.}
    \label{fig:results_pos}
\end{figure}

To analyse the query formulation in more detail for natural language, Figure \ref{fig:results_pos} presents the distribution of the eight most frequent grammatical types of words (parts of speech). While keyword searches should mainly consist of adjectives and nouns, grammatically complete sentences should contain a broader range of parts of speech. The results show that across conditions, the distribution and share of POS-tags changes. 

Vagueness is an inherent characteristic of natural language. In our experiment, we found significantly more vague words mentioned in I4 after the follow-up question (M = 3.3, STD = 2.7) than in either of the non-chatbot conditions, as Figure \ref{fig:boxplots}c shows. Looking at the percentage of vague words in key facts, no differences between conditions can be found (p = .972). Thus, the increase in vague words in I4 can be attributed to an increase in query length; key facts are not described more vaguely in any condition. The sentence completeness, however, correlates moderately positively with query vagueness (r = .50). Vagueness was more often used in the context of natural language, e.g.:
\begin{quote}
\centering
    \textit{``I'd like something \textbf{lightweight}, \textbf{easy} to use with a \textbf{long} battery life''} 
\end{quote}
Our results furthermore indicate that more domain knowledge reduces the number of vague words in a query (r = -.19).

\subsection{Perception of Participants}
After issuing the query, participants answered open questions about their experience with the search interface. The most striking observation was the repeated comparison with known search engines. 52\% of the participants mentioned in some way that they are not used to natural language input for search engines:
\begin{quote}
\centering
    \textit{``I wouldn't normally write in full sentences so it didn't feel natural.''} 
\end{quote}
Participants described the placeholder text in the query as a major signal for natural language, since it was a partial grammatical sentence. Participants furthermore suggested to use choice questions (``A or B?'') during a dialogue, react specifically to the initial query to show understanding, allowing voice input, and allowing to rank the mentioned attributes.

\section{Discussion}
In this study, we found that the chatbot interface elicits longer queries with a significantly higher number of key facts, especially after posing a follow-up question. Furthermore, the diversity of attributes was greater for the speech bubble and chatbot interface, as compared to both search bar interfaces. These findings suggest that the interface design has a major influence on the amount of information a user reveals in the initial query (\textbf{RQ 1}). We also found that longer queries contain more key facts and that participants with greater domain knowledge write longer queries. However, there is no correlation between the number of key facts and domain knowledge. This means that other factors besides domain knowledge drive the formulation of long queries with higher numbers of key facts. One possible driving factor for more key facts could be natural language.

While vagueness is a characteristic of natural language, we did not find evidence that people are more inclined to use vague language in our dialogue-inspired conditions. It could be that participants did not use vague words for convenience, but rather due to a lack of domain knowledge. We see that when domain knowledge increases, vagueness decreases. The lack of knowledge is not compensated by the interface, which results in equally vague formulations across conditions. If participants want a ``fast'' laptop, but do not know what facet this attribute contains, they will not know it regardless of the interface. The results, however, show a clear trend towards partial or full grammatical sentences in the chatbot interface, and the more sentence-like a query, the more key facts are mentioned. Together with the observed change in usage of parts of speech, we conclude that the queries of the chatbot interfaces are formulated in a more natural way than the queries in the search bar conditions (\textbf{RQ 2}).

Overall, the results demonstrate two prospects. (1) It is possible to influence and steer the formulation of queries towards a more natural language. (2) It is possible to stimulate the user of a product search system to reveal more (and more diverse) information about the desired product already in the initial query, but even more so with a generic follow-up question -- which is in line with the results reported in \cite{aliannejadi2019asking}. When stimulating users to reveal more information on the search target, search systems need to be equipped with appropriate functionality to process such long and diverse queries.


\section{Conclusion}
This study set out to investigate whether the interface design of a search engine can influence the amount of information revealed about a user's information need. We designed and executed an online user study (N=139) to investigate the query formulation for four search interfaces with varying conversational elements. 
Our key findings show that the interface design influences how and how much information users input into a search engine for the initial query. More precisely, conversation-like interfaces elicit more information and more natural language formulations at an initial stage of the information-seeking process than traditional search bar interfaces. 
This study reveals insights into the design of product search engines. For gaining a holistic image of the results across domains, follow-up research should investigate a broader variety of tasks to test the generalisability of the findings.

\begin{acks}
    This work was partly funded by the DFG, grant no. 388815326. We thank Alex Abas, Alexandra Bergen, Ali Khalife, Charlott Melcher, Annika Müller, Vanessa Sinanaj \& Saskia Zarges for their work on the interface designs and the study.
\end{acks}

\bibliographystyle{ACM-Reference-Format}
\bibliography{chiirsp03-bib}

\end{document}